\documentstyle[12pt]{article}

\newcommand{\p}{\bot}

\newcommand{\dd}{\partial}

\newcommand{\de}{\delta}
\newcommand{\De}{\Delta}
\newcommand{\om}{\omega}
\newcommand{\Om}{\Omega}
\newcommand{\e}{\varepsilon}
\newcommand{\f}{\varphi}
\newcommand{\ls}{\left(}

\newcommand{\rs}{\right)}

\newcommand{\g}{\gamma}

\newcommand{\be}{\beta}
\newcommand{\ta}{\tau}

\newcommand{\m}{\mu}
\newcommand{\s}{\sigma}

\newcommand{\et}{\eta}

\newcommand{\te}{\theta}
\newcommand{\sh}{{\rm sh}}
\newcommand{\ch}{{\rm ch}}
\newcommand{\tg}{{\rm tg}}
\newcommand{\ctg}{{\rm ctg}}

\newcounter{form}

\newcommand{\disn}[2]{$$\displaylines{\refstepcounter{form}
            \label{#1} \hfill #2}$$}
\newcommand{\no}{\hfill \phantom{(\theform)}\cr \hfill}
\newcommand{\nom}{\hfill (\theform) \cr}

\newcounter{punkt}
\makeatletter
\renewcommand{\section}{\@startsection{section}{1}{0pt}%
            {3.5ex plus 1ex minus .2ex}{2.3ex plus .2ex}{\bf}}
\long\def\@makecaption#1#2{%
   \vskip 10\p@
   \setbox\@tempboxa\hbox{#1. #2}%
   \ifdim \wd\@tempboxa >\hsize
       #1. #2\par
     \else
       \hbox to\hsize{\hfil\box\@tempboxa\hfil}%
   \fi}
\makeatother
\newcommand{\sect}[2]{\protect\refstepcounter{punkt}\protect\label{#1}
            \section*{$\protect\vphantom{a}$\hfill
            \arabic{punkt}.\hskip 2mm #2 \hfill $\protect\vphantom{a}$}}
\newcommand{\st}{\hfill $\protect\vphantom{a}$\protect\\
            $\protect\vphantom{a}$\hfill}

\begin{document}
\large

\title{$QED_2$ Light-Front Hamiltonian reproducing all \\
       orders of covariant  chiral perturbation theory}

\author{S.~A.~Paston\thanks{E-mail: Sergey.Paston@pobox.spbu.ru},
E.~V.~Prokhvatilov\thanks{E-mail: Evgeni.Prokhvat@pobox.spbu.ru},
V.~A.~Franke\thanks{E-mail: franke@snoopy.phys.spbu.ru}\\
St.-Peterspurg State University, Russia}

\date{October 15, 1999}

\maketitle

\begin{abstract}
Light-Front (LF) Hamiltonian for QED in (1+1)-dimensions is constructed
using the boson form of this model with additional Pauli-Villars type
ultraviolet regularization. Perturbation theory, generated by this LF
Hamiltonian, is proved to be equivalent to usual covariant chiral
perturbation theory. The obtained LF Hamiltonian depends explicitly on
chiral condensate parameters which enter in a form of some coupling
constants.
\end{abstract}

\newpage
\sect{vved}{Introduction}

Hamiltonian approach to Quantum Field Theory in Light-Front  (LF)
coordinates \cite{dir}
 \hbox{$x^{\pm}={1\over\sqrt{2}}(x^0\pm x^3)$},
 \hbox{$x^{\p}=\{x^1,x^2\}$}, with $x^+$
playing the role of time, is one  of  nonperturbative  approaches
which can be used in attempts to solve strong  coupling  problems
\cite{vils}.  It  has  the  advantage  of  having  simple   vacuum
state
description, because the physical vacuum is described on the LF
as the lowest eigenstate of LF momentum operator $P_-\ge 0$, and
this vacuum coincides with bare perturbative one on the LF.

    However the specific LF  singularities  at  zero  LF  momenta
($k_-\to 0$), being regularized via cutoff $|k_-|\ge\e >0$ (which
breaks Lorentz  and  gauge  symmetries),  can  be  the  cause  of
noncomplete equivalence  between  LF  theory  and  it's  original
formulation in Lorentz coordinates.  The  nonequivalence  can  be
found even when one compares LF perturbation  theory  with  usual
covariant one \cite{bur,tmf1}. One needs to add new  counterterms
to  canonical  Hamiltonian    to    restore    the    equivalence
\cite{bur,tmf1}. A general method to find these  counterterms  to
all orders in perturbation theory was described  in  \cite{tmf1}.
For gauge theories and, in  particular,  for  QCD,  in  axial  LF
gauge, $A_-=0$, this method gave an infinite number  of  possible
new counterterms \cite{tmf1}. To avoid this difficulty one needs
to use other, more complicated regularization scheme \cite{tmf2}.

   However  in  (1+1)-dimensional  space-time  one  can  use  the
bosonization  method  \cite{boz1,boz2,boz3,boz4}  to  treat   the
theory  analogously  to  scalar  field  theories.  These  bosonic
theories have essentially nonpolinomial form of  the  interaction
Hamiltonian. We show in the present paper how the method of paper
\cite{tmf1} can be applied to these theories. For  simplicity  we
use the example of $QED_2$. 
In this case the bosonization  method results in field theory like
Sine-Gordon model. As it is known, the Sine-Gordon model Lagrangian 
contains the interaction term in a form of $\cos(\be u)$, and there is 
special case $\be=\sqrt{4\pi}$, at which ultraviolet (UV)
behaviour of the theory  becomes worse. Exactly this case arises
in boson form of $QED_2$, and it is considered in this paper.
For a case $\be<\sqrt{4\pi}$  similar problem was investigated
in the paper \cite{burs}.
We construct LF Hamiltonian for $QED_2$ model (in boson form)  using
chiral perturbation  theory  to  all
orders.  This  Hamiltonian  depends  on   fermionic    condensate
parameters which enter the Hamiltonian like  coupling  constants.
These  parameters depend essentially on UV
regularization parameter (which is  introduced  via  Pauli-Villars
type regularization scheme) and can become infinite in the limit of
removed regularization cutoff.
The appearence of this divergency and  the necessity of UV regularization
in 
the model under consideration correspond to special case 
$\be=\sqrt{4\pi}$ in Sine-Gordon type models.

The obtained LF Hamiltonian can be applied to the calculation  of
mass spectrum using chiral perturbation theory. It can be checked
that results coincide with already known ones in  the  2nd  order
\cite{adam}. Moreover, one  can  apply  this  LF  Hamiltonian  in
nonperturbative    calculations    using    the    DLCQ    method
\cite{ann,paul}.

\sect{rasx}{The analysis of ultraviolet divergences}

  Let  us  start    with    the    boson    form    of    $QED_2$
\cite{boz1,boz2,boz3,boz4,boz}    in    Lorentz       coordinates
$x^{\m}=\{x^0,x^1\}$. It can be described by Lagrangian density
 \disn{1}{
L=\frac{1}{8\pi}\ls \dd_{\m}\f\dd^{\m}\f-m^2\f^2\rs-
\g\ls \cos\te-:\cos(\f+\te):\rs,
\nom}
where $\f(x)$ is scalar  field  originated  from  UV-renormalized
fermion  currents;  the  $\te$   --    parameter    characterizes
"instanton"  vacuum   \cite{boz3,adam};    $m=e/\sqrt{\pi}$    is
Schwinger  boson  mass,  $e$   is    the    coupling    constant;
$\g=Mme^c/(2\pi)$, $M$ is fermion mass and $c=0.577216$ is  Euler
constant;  normal  ordering  symbol,  $::$,  corresponds
to the decomposition of the quantized field  in  creation  and
annihilation operators in the interaction picture, the zero  mode
of the field is excluded,  $\int  dx^1\f(x)=0$  (that's  why  the
action doesn't include the term linear in $\f$).

Let  us  consider  the  structure  of  Feynman  diagrams  of  the
perturbation theory with respect to (w.~r.~t.) $\g$. The vertices
with $j$ entering lines give the factors
 \disn{dt3}{
i^{j+1}\g c_j(\te),\quad j\ge 2,
\nom}
where
 \disn{3}{
c_j(\te)=
\cases{\cos\te, & for even $j$ \cr
i \sin\te, & for odd $j$ \cr}.
\nom}
Propagators can be written as
 \disn{4}{
\De(k)=\frac{i}{\pi}\,\frac{1}{k^2-m^2+i0},\qquad
\De(x)=\int\! dk\, e^{ikx}\De(k),
\nom}
where $dk=dk_0dk_1$, $kx=k_0x^0+k_1x^1$.

First of all we investigate whether the  perturbation  theory  in
$\g$ is UV-finite. The interaction Hamiltonian is nonpolynomial in
the field $\f$, and one has infinite number of  diagrams  at  any
finite order in $\g$. One can see that all separate diagrams  are
finite (possible logarithmically  divergent  diagrams,  shown  in
fig.~\ref{r1}, are excluded owing to normal ordering).
 \begin{figure}[ht]
\begin{picture}(150,95)
\put(195,85){\special{em:graph fig1.bmp}}
\end{picture}
\caption{Logarithmically divergent diagram excluded by normal
ordering.}
\label{r1}
\end{figure}

Nevertheless infinite sum of these diagrams in a given  order  in
$\g$ can diverge, and this divergence is UV, because  it  can  be
removed by UV-regularization (as  it  is  shown  below).  Let  us
consider, for example, the sum $s_{ln}$
 \disn{4a}{
s_{ln}=\sum_{m=2}^{\infty}D_{lmn},\qquad s_{ln}=s_{nl}
\nom}
of diagrams $D_{lmn}$ shown in fig.~\ref{r2}.

 \begin{figure}[ht]
\begin{picture}(150,80)
\put(110,78){\special{em:graph fig2.bmp}}
\end{picture}
\caption{The  diagram  $D_{lmn}$  with  $l+n$  external  and  $m$
internal lines; $l$ and $n$ lines are connected to the  left  and
right vertex,  accordingly;  $p$  is  the  total  momentum  going
through the diagram.}
\label{r2}
\end{figure}
This sum is
 \disn{2}{
s_{ln}=
\frac{\g^2}{l!n!(1+\de_{ln})}
\sum_{m=2}^{\infty}\frac{1}{m!}c_{l+m}(\te)i^{l+m+1}
c_{n+m}(\te)i^{n+m+1}\times\no
\times\int \prod_{j=1}^m dk_j\,
(2\pi)^2\; \de\Bigl(\sum_{j=1}^m k_j-p\Bigr)\,\prod_{j=1}^m \De (k_j),
\nom}
where $p$ is total momentum going through the diagram, $\de_{ln}$
is Kronecker symbol. "Symmetry" coefficients of the diagrams  are
taken  so  that  the  sum  of  expressions  $s_{ln}$  with    all
transpositions of external momenta gave the contribution to Green
function. We obtain
 \disn{5}{
s_{ln}=\frac{\g^2}{l!n!(1+\de_{ln})}
\sum_{m=2}^{\infty}\frac{1}{m!}c_{l+m}(\te)
c_{n+m}(\te)i^{l+n+2m+2}\int\! dx\, e^{-ipx} \De(x)^m=\no
=-\frac{\g^2i^{l+n}}{l!n!(1+\de_{ln})}
\int\! dx\, e^{-ipx}\times\hskip 67mm\no
\times \ls c_{l+1}(\te)c_{n+1}(\te)\ls \De(x)-\sh(\De(x))\rs+
c_l(\te)c_n(\te)\ls \ch(\De(x))-1\rs\rs,
\nom}
where the eq.~(\ref{3}) was taken into account. At $x\to 0$
 \disn{6}{
\De(x)\sim \ln\frac{1}{x^2},
\nom}
so that we get the integral over $x$ logarithmically divergent.

Let us show that, in spite  of  this,  the  sum  of  all  diagrams
$s_{ln}$, contributing to Green functions, is finite. Indeed  the
divergent part of (\ref{5}) is proportional to
 \disn{7}{
s_{ln}^{\infty}=-\frac{\g^2i^{l+n}}{2l!n!(1+\de_{ln})}
\ls c_l(\te)c_n(\te)-c_{l+1}(\te)c_{n+1}(\te)\rs.
\nom}
As it follows from definitions (\ref{3}), the quantity  (\ref{7})
is
 \disn{8}{
s_{ln}^{\infty}=-\frac{\g^2i^{l+n}}{4l!n!(1+\de_{ln})}
\ls (-1)^l+(-1)^n\rs.
\nom}
The Green function includes the sum
 \disn{9}{
\sum_{0\le l\le j/2}s_{l,j-l},
\nom}
the divergent part of it is
 \disn{10}{
\sum_{0\le l\le j/2} s_{l,j-l}^{\infty}=
\frac{1}{2}\sum_{l=0}^j s_{l,j-l}^{\infty}(1+\de_{l,j-l})=\no=
-\frac{\g^2i^j}{8j!}
\ls \sum_{l=0}^j \frac{j!}{l!(j-l)!} (-1)^l+
\sum_{l=0}^j \frac{j!}{l!(j-l)!} (-1)^{j-l}\rs=\no=
-\frac{\g^2i^j}{8j!}
\ls (-1+1)^j+(1-1)^j\rs=0,
\nom}
as it is stated above.

We shall suppose 
that UV-finiteness of Green functions holds to any order  in  $\g$  in
this model. But we must consider always only sums like (\ref{9}),
but not separate  $s_{ln}$.  As  it  will  be  explained    in
Sect.~\ref{postr}, this fact makes  impossible  the  perturbative
construction of LF Hamiltonian, by the method of \cite{tmf1},  if
we don't introduce proper UV regularization.

\sect{postr}{The perturbative construction of the Hamiltonian}

Let us try to construct  a  LF  Hamiltonian  regularized  by  the
cutoff in LF momenta, $|k_-|\ge\e >0$, (i.~e. there are no  modes
with $|k_-|<\e$ in the Fourier decomposition of the field  $\f$),
and to generate a perturbation theory, equivalent  to  the  usual
covariant one in the limit $\e\to 0$. Let us start from canonical
LF Hamiltonian which follows from the Lagrangian  (\ref{1}).  The
interaction part of this Hamiltonian has the form:
 \disn{29}{
H_I^{can}=\g\ls \cos\te -:\cos(\f+\te):\rs=\no
=\g\cos\te\ls 1-:\cos\f:\rs+\g\sin\te:\sin\f:.
\nom}
It can be shown  \cite{har,lay}  that  noncovariant  perturbation
theory, obtained with LF Hamiltonian,  can  be  transformed  into
equivalent  covariant  perturbation  theory  via  resummation  of
diagrams, but in this theory  the  integrations  over  $k_-$  are
limited by cutoff $|k_-|\ge\e$, and  one  should  also  integrate
firstly over $k_+$, then over $k_-$. The $\e\to 0$ limit of  such
LF calculation of the diagram we call LF diagram. To construct LF
Hamiltonian  one  has  to  find  all  necessary  counterterms  to
canonical LF Hamiltonian which compensate the differences between
LF and covariant diagrams.

The method of estimating these differences that  allows  to  find
all  diagrams,  giving  nonzero  difference,  is  described    in
\cite{tmf1}. For its application  the  requirement  $\om_+<0$  is
necessary, where $\om_+$ is the index  of  UV-divergence  of  the
diagram in $k_+$. This requirement means that  Feynman  integrals
are to be UV-convergent in $k_+$. It is easy to see, that for the
considered theory this requirement is fulfiled.

Let us describe briefly the method of \cite{tmf1}. Consider,  for
example,  an  arbitrary  1-loop  Feynman  diagram  with  external
entering momenta $p^i_{\mu}$, $i=1,2,\dots$.  The  loop  momentum
$k_-$ is bounded by cutoff  conditions  $|k_--\sum  p^i_-|\ge\e$,
steming from restriction on the propagator. On the  other  side,
analogous covariant diagram contains  the  integration  over  all
$k$. Therefore the difference between these diagrams can be found
as the sum of integrals over the bands $ |k_- - \sum p_-^i|<\e$.

Let us estimate one of these "$\e$"-band integrals. We shift  the
variable $k_-$ in this  integral  so  that  $|k_-|<\e$.  Then  we
change the scale:
 \disn{11a}{
k_-\longrightarrow \e k_-,\qquad k_+\longrightarrow \frac{1}{\e} k_+.
\nom}
that makes the integration interval independent  of  $\e$,  while
keeps Lorentz invariant products  like  $k_+k_-$  or  $dk_+dk_-$.
Propagators, corresponding to internal lines  whose  momenta  are
outside of the $\e$-band (owing to external momentum $p_-$, going
through the line) change as follows:
 \disn{11b}{
\frac{i}{\pi}\,\frac{1}{2(k_++p_+)(k_-+p_-)-m^2+i0}\longrightarrow\no
\longrightarrow
\frac{i}{\pi}\,\frac{1}{2(\frac{1}{\e}k_++p_+)(\e k_-+p_-)-m^2+i0}
\mathrel{\mathop{\approx}\limits_{\e\to 0}}
\frac{i}{\pi}\,\frac{\e}{2k_+p_-}.
\nom}
In the paper \cite{tmf1} we used denotations for the  lines  with
momenta outside and inside of the $\e$-band. First one was called
$\Pi$-line  and  the  last  one,  $\e$-line.  It  follows    from
eq.~(\ref{11b}) that every $\Pi$-line gives  a  factor  of  order
$O(\e)$ while every $\e$-line gives a  factor  of  order  $O(1)$.
Therefore the integral over the band is zero in $\e\to  0$  limit
if at least one of $\Pi$-lines is present in the diagram.

Similar analysis can be made for an arbitrary  many-loop  Feynman
diagram. The difference between LF and covariant  calculation  of
this diagram can be estimated again by considering  all  possible
configurations  of  $\Pi$-  and  $\e$-  lines  in  the    diagram
\cite{tmf1}. It was shown in the paper \cite{tmf1}  (for  a  wide
class of field theories) that each of these configurations can be
estimated in $\e$ as having the  order  $O(\e^{\s})(1+O(\log\e))$
with
 \disn{11}{
\s=min(\ta,\om_--\om_+-\m+\et),
\nom}
where the minimum is to be taken w.r.t. all  subdiagrams  of  the
diagram at some configuration of $\Pi$- and $\e$-  lines  in  it;
$\omega_{\pm}$ are indices of UV-divergency  in  $k_{\pm}$  of  a
given subdiagram; $\mu$ is the index of  total  UV-divergency  in
$k_-$ of all $\Pi$-lines in the subdiagram; $\tau$ is total power
of the $\e$ that arises, after the change $k_-\to \e k_-$ of loop
variables $k_-$,  from  numerators  of  all  propagators  of  the
diagram and from all volume elements in the integrals over $k_-$;
$\eta$ is  the  part  of  the  $\tau$  related  with  only  those
numerators and volume elements (used in the definition of $\tau$)
that are not present in the considered subdiagram. Let  us  apply
this general result to our scalar field theory.  All  propagators
have simple structure. Only possible contribution to  the  $\tau$
comes from the volume elements $dk_-$. Because we are  interested
only  in  the  difference  of  LF  and  covariant  diagrams,  any
configuration should contain at least one integration over  $k_-$
in the $\e$-band. Therefore, $\tau >0$ (and $\eta\ge 0$). Due  to
Lorentz-invariant form of diagrams in $k_+,k_-$ we have
 $\om_+-\om_- =0$. It follows from the expression (\ref{11b}) for
a $\Pi$-line propagator that the $\mu$  can  be  counted  as  the
number of $\Pi$-lines in the  subdiagram  taken  with  the  minus
sign. Therefore, one has $ -\mu + \eta >0$ (and,  hence,  $\s>0$)
if  at  least  one  of  $\Pi$-lines  is  present.   Thus,    only
configurations without $\Pi$-lines, i.~e. at  $\m=\et=\s=0$,  can
contribute to the difference between LF and  covariant  diagrams.
All these configurations are connected with  diagrams,  shown  in
Fig.~\ref{r3}, which are equal to zero in LF perturbation theory.
Therefore, all difference between LF and  covariant  perturbation
theory is related with the sum of these covariant diagrams.
 \begin{figure}[ht]
\begin{picture}(150,140)
\put(170,135){\special{em:graph fig3.bmp}}
\end{picture}
\caption{General form of diagrams giving the contribution to  the
difference between LF and covariant perturbation theories.}
\label{r3}
\end{figure}

However the conclusion made before was based on the estimation of
the difference between LF  and  covariant  calculation  for  each
separate diagram. To estimate this difference for infinite sums of
diagrams which are present in our model at  any  given  order  of
perturbation theory, one must  be  sure  that  infinite  sums  of
separate estimations converge uniformly in $\e$. But it is not so
in our scheme, due to divergency of some partial sums of diagrams
(like $s_{ln}$ that we considered above). That is why we  use  in
the following some additional  regularization  that  makes  these
partial sums finite.

\sect{pavi}{The construction of LF Hamiltonian \st
with Pauli-Villars type regularization}

We need a Lorentz invariant regularization that can  be  used  in
covariant and LF calculations simultaneously. We use Pauli-Villars
type one, modifying the Lagrangian (\ref{1}) in the following way:
 \disn{13}{
L=\frac{1}{8\pi}
\sum_{l=0,1}(-1)^l\ls \dd_{\m}\f_l\dd^{\m}\f_l-m_l^2\f_l^2\rs-
\g\ls \cos\te-:\cos(\f+\te):\rs,
\nom}
where $\f =\f_0 + \f_1$, $\f_0 $ being the original field of  the
mass $m_0 = m$, and $\f_1$ being  the  additional  (ghost)  field
with a large mass $m_1$ playing the role of UV cutoff.  Only  the
sum of propagators of $\f_0$ and $\f_1$ fields enter the  Feynman
diagrams. This sum defines regularized propagator in the form
 \disn{13a}{
\De(k)=\frac{i}{\pi}\,\frac{1}{k^2-m_0^2+i0}-
\frac{i}{\pi}\,\frac{1}{k^2-m_1^2+i0}=\no=
\frac{i}{\pi}\,\frac{m_0^2-m_1^2}{(k^2-m_0^2+i0)(k^2-m_1^2+i0)}.
\nom}

At finite $m_1$ the corresponding propagator $\De(x)$  is  finite
at $x=0$, and one avoids the divergency for sums  like  $s_{ln}$.
One  can  repeat  now  all  arguments  of  Sect.~\ref{postr}  and
conclude  that  the  difference  between   LF    and    covariant
perturbation  theories  exists  only  for  diagrams   shown    in
Fig.~\ref{r3}. And the difference between these covariant and  LF
diagrams coincides with covariant diagrams, because  LF  diagrams
are equal to zero. Therefore this difference does not  depend  on
$\e$.

The counterterms which must be added to canonical LF  Hamiltonian
should generate all  diagrams  shown  in  Fig.~\ref{r3}.  Let  us
denote such arbitrary diagram of order $n$, with $l$ external and
$m$ internal lines entering the same vertex, by  $R^{n(i)}_{lm}$,
where  the  index $i$ numerates different these diagrams at fixed
$n,l,m$. One  can  express  all  these  diagrams  in   terms   of
only   $R^{n(i)}_{0m}$   taking  into  account   vertex   factors
(\ref{dt3}) and symmetry coefficients of diagrams:
 \disn{14}{
R_{lm}^{n(i)}=
\cases{
R_{0m}^{n(i)}r^{n(i)}_m(-1)^{l/2}, &for even $l$ \cr
R_{0m}^{n(i)}r^{n(i)}_m(-1)^{(l-1)/2}(-\tg\te),
&for odd $l$ and even $m$ \cr
R_{0m}^{n(i)}r^{n(i)}_m(-1)^{(l+1)/2}(-\ctg\te),
&for odd $l$ and $m$ \cr},
\nom}
where $r^{n(i)}_m$  is  a  number  of  vertices  in  the  diagram
$R^{n(i)}_{0m}$ which are equivalent to the vertex, to which  the
external  momenta  are  joined  in  the  diagram  $R^{n(i)}_{lm}$
(including this vertex).

The counterterm generating the sum  of  diagrams  $R^{n(i)}_{lm}$
with all even $l$ is
 \disn{17}{
\sum_{k=1}^{\infty}R_{2k,m}^{n(i)}\frac{i}{(2k)!} :\f^{2k}:=
\sum_{k=1}^{\infty}R_{0m}^{n(i)}r^{n(i)}_m(-1)^k
\frac{i}{(2k)!} :\f^{2k}:=\no=
-ir^{n(i)}_mR_{0m}^{n(i)}\ls 1-:\cos\f:\rs.
\nom}
The  quantity  $R^{n(i)}_{0m}$  is  one  of    vacuum    diagrams
contributing  to  vacuum  energy  density.  Summing    expressions
(\ref{17})  for  all  different  $R^{n(i)}_{lm}$  with  the  same
$R^{n(i)}_{0m}$, and then summing over $i$ and $m$, we get
 \disn{18}{
H_1^n=-in\ls 1- :\cos\f:\rs\sum_{i,m}R_{0m}^{n(i)},
\nom}
because the external lines can be  joined  to  representative  of
each group of equivalent vertices (to one representative for each
group), and the  number  of  vertices  in  such  group  is  equal
$r^{n(i)}_m$. Summing also over all $n$ we  obtain  the  following
form of the counterterm generating all diagrams of  Fig.~\ref{r3}
with even number of external lines:
 \disn{18a}{
H_1=\sum_n H_1^n=\tilde C_1\ls 1- :\cos\f:\rs,\qquad
\tilde C_1=-i\sum_{i,m,n} nR_{0m}^{n(i)}.
\nom}
The coefficient $\tilde C_1$ depends on parameters  $\g$,  $\te$,
$m_0$, $m_1$ and can become infinite in $m_1\to\infty$ limit.

For  the odd $l$ we can find, that
 \disn{21}{
\sum_{k=0}^{\infty}R_{2k+1,m}^{n(i)}\frac{i}{(2k+1)!} :\f^{2k+1}:=\nom=
\sum_{k=0}^{\infty}R_{0m}^{n(i)}r^{n(i)}_m(-1)^k(-\tg\te)
\frac{i}{(2k+1)!} :\f^{2k+1}:
=-i\tg\te r^{n(i)}_mR_{0m}^{n(i)} :\sin\f:\hskip -24pt
\no}
for even $m$, and
 \disn{21a}{
\sum_{k=0}^{\infty}R_{2k+1,m}^{n(i)}\frac{i}{(2k+1)!} :\f^{2k+1}:=\nom=
\sum_{k=0}^{\infty}R_{0m}^{n(i)}r^{n(i)}_m
(-1)^{k+1}(-\ctg\te) \frac{i}{(2k+1)!} :\f^{2k+1}:=
i\ctg\te r^{n(i)}_mR_{0m}^{n(i)} :\sin\f:\hskip -27pt
\no}
for odd $m$. Every vertex in the diagram $R^{n(i)}_{0m}$ has  the
factor  $\cos\te$  if  the  number of it's legs is even, and  the
factor  $\sin\te$ if this number is odd. The factors $-\tg\te$ in
(\ref{21})   for   even   $m$,  or  $\ctg\te$  for  odd   $m$  in
(\ref{21a}), can   be    produced    by   the   action   of   the
derivative $\frac{\dd}{\dd\te}$ on vertex factors  $\cos\te$,  or
$\sin\te$,  accordingly. Summing at fixed $m$  expressions 
 (\ref{21})  and (\ref {21a}) for all  different  $R^{n(i)}_{lm}$
with  the    same   $R^{n(i)}_{0m}$  and  taking into account the
factors  $r^{n(i)}_m$,   we   get  the sum of $n$ terms, in  each
of   which  the    derivative  $\frac{\dd}{\dd\te}$ acts  on  its
own  factor, so  that  after summing over $i$ and $m$ we have
 \disn{22}{
H_2^n=i :\sin\f :\frac{\dd}{\dd\te}\sum_{i,m}R_{0m}^{n(i)}.
\nom}
Summing over all $n$ we obtain  the  counterterm  generating  all
diagrams of Fig.~\ref{r3} with odd number of external lines:
 \disn{22a}{
H_2=\sum_n H_2^n=\tilde C_2:\sin\f:,\qquad
\tilde C_2=i\frac{\dd}{\dd\te}\sum_{i,m,n} R_{0m}^{n(i)}.
\nom}
The coefficient $\tilde C_2$ depends on the  same  parameters  as
$\tilde C_1$.

Now we can write the  corrected  LF  interaction  Hamiltonian  as
follows:
 \disn{22b}{
H_I=H_I^{can}+H_1+H_2=C_1\ls 1-:\cos\f:\rs+C_2:\sin\f:\no
C_1=\g\cos\te+\tilde C_1,\quad C_2=\g\sin\te+\tilde C_2.
\nom}
The corrected LF  Hamiltonian  generates  the  theory,  which  is
perturbatively  equivalent  to  covariant    theory,    if    the
coefficients   $C_1$   and   $C_2$   depend    properly    on the
parameters $\g$ and $\te$ of the covariant theory.

Let us show that the coefficients $C_1$, $C_2$ can be written  in
the form of some condensate parameters of the field quantized  at
$x^0=0$  in  Lorentz  coordinates.  Firstly,  let  us  write  the
eq.~(\ref{18a}) in the form
 \disn{22c}{
\tilde C_1=-i\g\frac{\dd}{\dd\g}\sum_{i,m,n} R_{0m}^{n(i)}=
-i\g\frac{\dd}{\dd\g}\tilde G_0,
\nom}
where $\tilde G_0$ is  the  density  of  connected  vacuum  Green
function, i.~e.
 \disn{24}{
\tilde G_0=\frac{1}{V}\ln G_0,
\nom}
where the $V$ is the space-time volume, and the $G_0$ is vacuum  Green
function:
 \disn{25}{
G_0=\langle 0|{\rm T}\exp\ls iS_I\rs|0\rangle,
\nom}
with  the  interaction  part  of  the  action  defined  in    the
interaction picture by the expression
 \disn{26}{
S_I=\int\! dx\, \g\ls :\cos(\f+\te):-\cos\te\rs
\nom}
in Lorentz coordinates. Analogously, the eq.~(\ref{22a})  can  be
written in the form
 \disn{22d}{
\tilde C_2=i\frac{\dd}{\dd\te}\tilde G_0.
\nom}
Hence we get
 \disn{27}{
C_1=\g\cos\te-i\frac{\g}{VG_0}
\langle 0|{\rm T}\int\! dx\, i\ls:\cos(\f+\te):-\cos\te\rs e^{iS_I}|0\rangle
=\no
=\g\cos\te+\g\frac{
\langle 0|{\rm T} \ls:\cos(\f+\te):-\cos\te\rs e^{iS_I}|0\rangle}{
\langle 0|{\rm T} e^{iS_I}|0\rangle}=\no=
\g\cos\te+\g\langle \Om|\ls:\cos(\f+\te):-\cos\te\rs |\Om\rangle=
\g\langle \Om|:\cos(\f+\te):|\Om\rangle,
\nom}
and
 \disn{28}{
C_2=\g\sin\te+i\frac{1}{VG_0}
\langle 0|{\rm T}\int\! dx\, i\g\ls-:\sin(\f+\te):+\sin\te\rs
e^{iS_I}|0\rangle=\no
=\g\sin\te+\g\frac{
\langle 0|{\rm T} \ls:\sin(\f+\te):-\sin\te\rs e^{iS_I}|0\rangle}{
\langle 0|{\rm T} e^{iS_I}|0\rangle}=\no=
\g\sin\te+\g\langle \Om|\ls:\sin(\f+\te):-\sin\te\rs |\Om\rangle=
\g\langle \Om|:\sin(\f+\te):|\Om\rangle,
\nom}
where  the  $|\Om\rangle$  is  physical  vacuum  state.  Now  the
interaction Hamiltonian (\ref{22b}) can be rewritten in the form
 \disn{30}{
H_I=\g\ls 1-:\cos\f:\rs\langle \Om|:\cos(\f+\te):|\Om\rangle+\no+
\g:\sin\f:\langle \Om|:\sin(\f+\te):|\Om\rangle.
\nom}

Notice that the normal ordering inside of vacuum matrix  elements
corresponds to bare (perturbative) vacuum of usual formulation in
Lorentz coordinates. It follows from the  bosonization  procedure
that these vacuum  condensate  parameters  are  equal  (up  to  a
constant, included  in  the  $\gamma$)  to  fermionic  condensate
parameters        $\langle\bar\Psi\Psi\rangle$                and
$\langle\bar\Psi\g^5\Psi\rangle$, which require UV-regularization
\cite{adam}, achieved now  by  introducing  the  cutoff  mass
$m_1$.

   Let us notice also that the  formula  (\ref{30})  for  the  LF
Hamiltonian can be  directly  obtained  via  the  method  of  the
limiting  transition  to  LF  Hamiltonians  starting  from  usual
Hamiltonian formulation \cite{pred}. The  approximation  used  in
\cite{pred} becomes true owing to the analysis of the  difference
between LF and covariant perturbation theories diagrams,  carried
out above, so that this method also can be applied.

One can  use  the  obtained  LF  Hamiltonian  in  nonperturbative
calculations of the mass spectrum, that can be  done  on  the  LF
numerically by  DLCQ  method  \cite{ann,paul},  and  compare  the
results with those obtained in usual formulations of $QED_2$.  In
these calculations the cutoff in momenta $k_-$  must  be  removed
before the removing UV cutoff.

\vskip 5mm
\centerline{\bf Acknowledgements}
\nopagebreak
\vskip 2mm

The author (E.~V.~P.) is  grateful  to  the  DAAD  for  financial
support. The present work was supported also in part (for one  of
authors S.~A.~P.) by the INTAS Grant 96-0457 within the  research
programm of the International Center for Fundamental  Physics  in
Moscow. The authors are grateful to profs. F.~Lenz  and  M.~Thies
for useful discussions at Erlangen-Nuernberg University.

\end{document}